\documentclass[a4paper,10pt]{article}
\pdfoutput=1 

\usepackage{jinstpub} 

\usepackage{xspace}
\usepackage{lineno,hyperref}

\newcommand{\sFGD}{sFGD\xspace}

\title{\boldmath A $4\pi$ time-of-flight detector for the ND280/T2K upgrade}

\author[a,1]{A.\,Korzenev\note{Current address: Joint Institute for Nuclear Research, Dubna, Russia.},}
\author[b]{F.\,Barao,}
\author[a]{S.\,Bordoni,}
\author[c]{D.\,Breton,}
\author[a]{F.\,Cadoux,}
\author[a]{Y.\,Favre,}
\author[d]{M.\,Khabibullin,}
\author[d]{A.\,Khotyantsev,}
\author[d,e,f]{Y.\,Kudenko,}
\author[g]{T.\,Lux,}
\author[c]{J.\,Maalmi,}
\author[a,2]{P.\,Mermod%
\note{Deceased in the summer of 2020.},}
\author[d]{O.\,Mineev,}
\author[a]{F.\,Sanchez}

\affiliation[a]{DPNC, Section de Physique, Universit\'{e} de Gen\`{e}ve, Geneva, Switzerland}
\affiliation[b]{Laboratory of Instrumentation and Experimental Particles Physics,  Lisboa and  Coimbra, Portugal}
\affiliation[c]{Laboratoire de L'acc\'{e}l\'{e}rateur Lin\'{e}aire from CNRS/IN2P3, Orsay, France}
\affiliation[d]{Institute for Nuclear Research of the Russian Academy of Sciences, Moscow, Russia}
\affiliation[e]{Moscow Institute of Physics and Technology (MIPT), Moscow, Russia}
\affiliation[f]{National Research Nuclear University (MEPhI), Moscow, Russia}
\affiliation[g]{Institut de F\'isica d'Altes Energies (IFAE) - The Barcelona Institute of Science and Technology (BIST), Campus UAB,  Bellaterra (Barcelona), Spain}

\emailAdd{alexander.korzenev@cern.ch}

\abstract{ND280 is a near detector of the T2K experiment which is located in the J-PARC accelerator complex in Japan.
After a decade of fruitful data-taking, ND280 is scheduled for upgrade.
The time-of-flight (ToF) detector, which is described in this article, is one of three new detectors
that will be installed in the basket of ND280. 
The ToF detector has a modular structure. 
Each module represents an array of 20 plastic scintillator bars which are stacked in a plane
of $2.4 \times 2.2 $~m$^2$ area.
Six modules of similar construction will be assembled in a cube, thus providing 
an almost $4\pi$ enclosure for an active neutrino target and two TPCs.
The light emitted by scintillator is absorbed by arrays of large-area silicon photo-multipliers (SiPMs)
which are attached to both ends of every bar.
The readout of SiPMs, shaping and analog sum of individual SiPM signals within the array
are performed by a  discrete circuit amplifier. 
An average time resolution of about 0.14~ns is achieved for a single bar when measured with cosmic muons.
The detector will be installed in the basket of ND280, where
it will be used to veto particle originating outside the neutrino target,
 improve the particle identification and  provide a cosmic trigger for calibration of detectors which are enclosed inside it.}


\begin{document}

\maketitle
\flushbottom

\section{Introduction}

\begin{figure*}[t]
\includegraphics[width=0.6\textwidth]{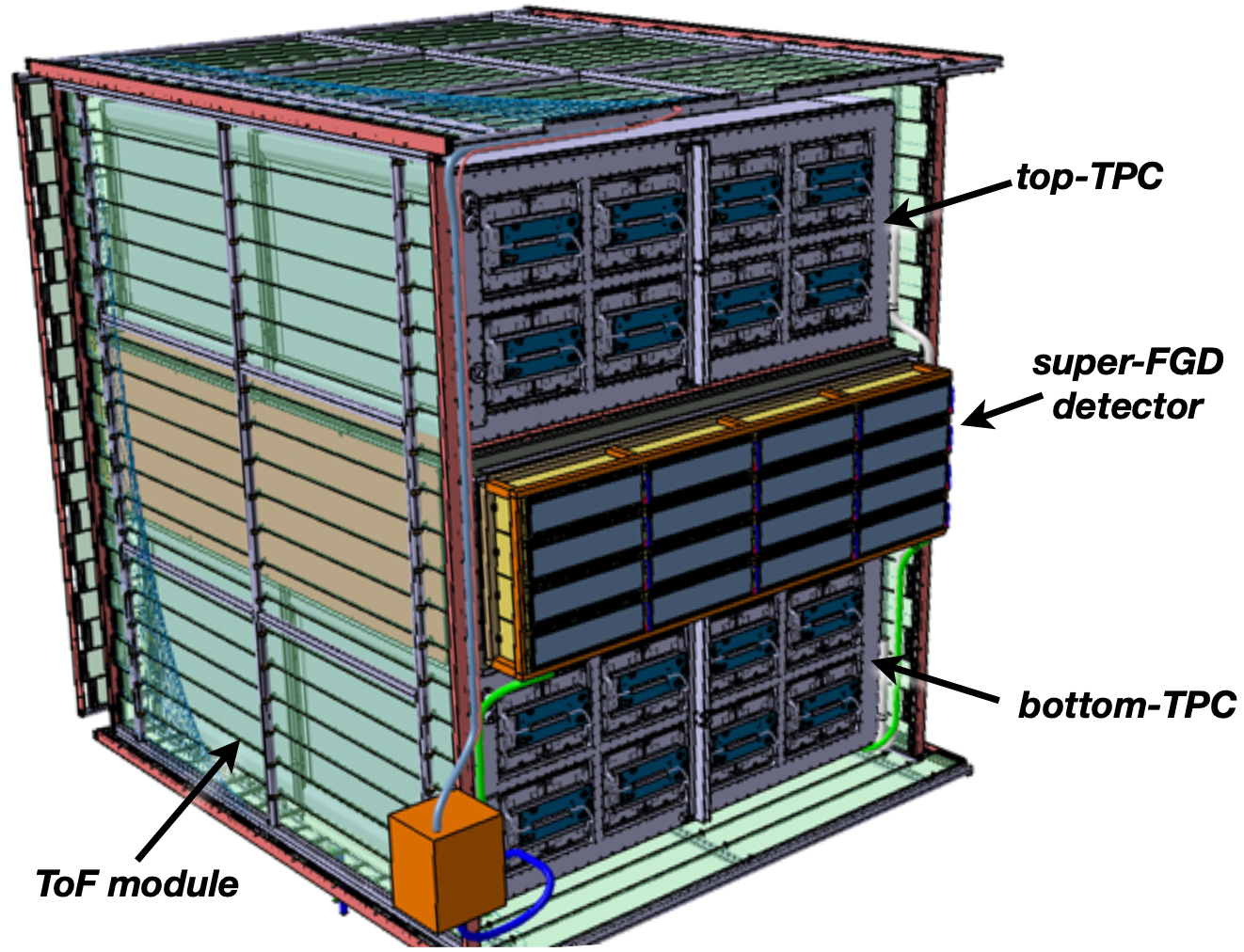}
\hfill
\includegraphics[width=0.36\textwidth]{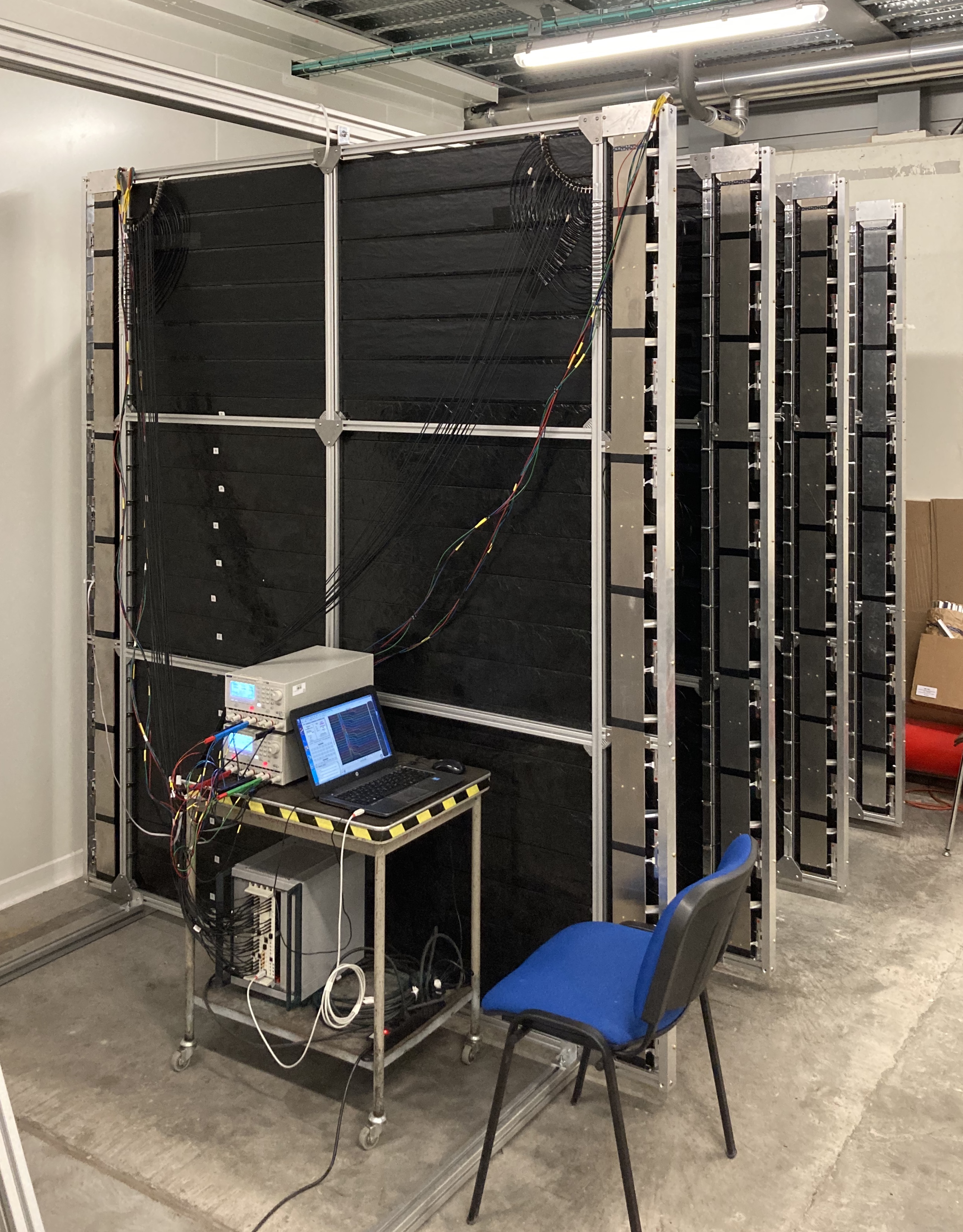}
\caption{ {\it Left:} six modules of ToF assembled in a cube  covering the \sFGD and two TPC detectors.
The front ToF module is not drawn in order to see the interior.
{\it Right:}  photo of four ToF modules during the test with cosmics in the CERN neutrino platform building.
}
  \label{fig_photo}
\end{figure*}

An improvement of characteristics of the ND280 detector is one of the goals of the ongoing T2K upgrade
which is a part of the second phase of the T2K experiment \cite{T2K:2019bbb}.
For this purpose, an active neutrino target (\sFGD), consisting of small scintillator cubes \cite{Mineev:2019dpe,Blondel:2020hml},  
and two TPCs, which will cover large angles \cite{Attie:2019hua}, will be installed in the basket of ND280.
The time-of-flight system (ToF) will surround these detectors providing a close to 4$\pi$ coverage as shown in Fig.\,\ref{fig_photo} (left).
ToF will serve to determine the  direction of charged particles (inward or outward with respect to the target), thereby
rejecting those originating from interactions in materials
 outside  the target volume.
The corresponding uncertainty is currently one of the main contributors to the systematic uncertainty  the ND280  physical results  \cite{Abe:2018uhf}.
A time resolution of 0.5\,ns is required to unambiguously determine the direction of particle flight.
ToF can also be used to improve particle identification
 if one can achieve the time-of-flight resolution better than 0.2\,ns.
 Although this identification is mostly ensured by $dE/dx$ measurements in the TPCs, 
discrimination between muons and electrons or 
between protons and positrons is not possible in some momentum ranges due to similar energy loss. 
In addition, the fact that ToF encloses the \sFGD and TPC detectors makes it convenient for triggering on cosmic muons.
The ToF will also ensure a precision timing reference for detectors enclosed inside it.

The paper is organized as follows.
In this section we introduce the ToF detector concept. 
Details on the amplifier board are given in Sec.\,\ref{sec:amplifier}. 
Results of the timing characteristic measurements for a single ToF bar are presented in Sec.\,\ref{sec:aingle_bar}.
The paper concludes with Sec.\,\ref{sec:conclusions}.


\subsection{Detector concept}

Plastic scintillators, due to their fast photon emission properties,  are widely employed for neutrino detectors that are located on the surface of the earth.
Neutrino beams are bunched with a duration of the order of microseconds, therefore the cosmic ray background is 
filtered out during analysis using timing information.

There are two types of plastic scintillators commonly used in experiments:
an extruded plastic with wavelength shifting (WLS) for readout and a cast scintillator with direct light readout. 
We briefly compare their relevant parameters.
A typical number of optical photons released isotropically by a minimum-ionising particle crossing 
a 1\,cm plastic scintillator is $\sim$10$^4$.
The resulting number of detected photons  depends on a distance of propagation and photo-detection efficiencies. 
In the case of extruded plastic, the scintillator is usually shaped as a bar
with a central hole \cite{Amaudruz:2012agx}   or a surface groove \cite{Aoki:2012mf}  for a WLS fiber.
A small trapping angle for transmitted photons in the fiber strongly limits the absolute amount of light collected.
Typically, several 10's of photons are detected.
The time spread 
due to re-emission in the WLS material
introduces an extra uncertainty for the time measurement.
That makes the overall time resolution to be around 1\,ns.

The time resolution of 0.5~ns which is necessary for determining the direction of flight in ND280,
would require several layers of plastic scintillator described above. 
A better value of the resolution assumes an even larger number of layers which makes this approach inconvenient.
Given a very limited space inside the ND280 basket, the solution with a cast plastic scintillator
that propagates near-UV photons of the primary fluor has been taken as a baseline for the ToF detector. 
The bulk attenuation length of several meters and 
very low light losses in successive internal reflections at the surface of the bar
eliminate the need for WLS and
allow the number of photons detected at the bar ends  to be of the order of hundreds.
The combination of high light yield and fast response time of the scintillator 
makes possible a reduction in the uncertainty typically down to the
0.1~ns level for bars with a length of 2~m. 
This improvement of the time resolution by an order of magnitude compared to the case of extruded plastic 
would require, however, a photo-sensor with a much larger surface coverage.

Traditionally, vacuum photomultiplier tubes (PMTs) are utilised to convert the light to an electronic signal.
However, due to the  limited space and 0.2\,T magnetic field inside the ND280, this option was not possible.
A~readout by arrays of large-area silicon photomultipliers (SiPMs) was chosen as a baseline.
This approach was already tested earlier for the readout of a single plastic scintillator bar~\cite{JINST,Korzenev:2019kud} and WLS optical modules~\cite{Ehlert:2018pke}.
Nowadays, large-area SiPMs have appeared on the market at relatively low cost and 
offer several advantages over PMTs: 
magnetic field tolerance, a much smaller volume and footprint allowing a compact design for bars without light-guides,
low operation voltage and high photon detection efficiency.
An increased sensitivity of SiPMs to the longer-$\lambda$ of the optical spectrum, which is least attenuated inside the plastic, favours the use of longer bars.


\subsection{Assembly of the ToF system}

The ToF system consists of six modules of similar construction, which are assembled in a cube
as shown in Fig.\,\ref{fig_photo} (left).
Each module is a set of 20 plastic scintillator bars which are arranged in a plane
with a total active area of 
$5.4 $~m$^2$. 
In view of the limited space inside the basket, demands for robustness of the construction\footnote{
The module must have a shock resistance of $0.5g$ in the transverse direction to the plane, according to the J-PARC safety regulations. 
}
and similar dimensions for all modules,
the configuration with staggered bars was disfavoured. The planar configuration makes the detector cheaper and easier to assemble. 
Bars are therefore positioned in a plane with a gap of 1.5~mm between the side faces. The gap is necessary to include steel brackets which keep the bars attached to the outer aluminum frame.
Two bars of the bottom module are removed to leave space for trays which will guide service cables of \sFGD and TPCs outside the basket.

The size of each  bar is $220 \times 12 \times 1$~cm$^3$. 
The material is a plastic scintillator EJ-200 with a scintillation rise time of 0.9 ns, 
a decay time of 2.1~ns and an attenuation length of 380~cm \cite{SCIONIX}. 
The wavelength emission spectrum peaks at  425~nm perfectly matching the SiPM spectral response. 
The bars were wrapped in aluminium foils and a black plastic stretch film on top to ensure opacity.

\begin{figure}[t]
\includegraphics[width=0.49\columnwidth]{Fig2_1}
\hfill
\includegraphics[width=0.49\columnwidth]{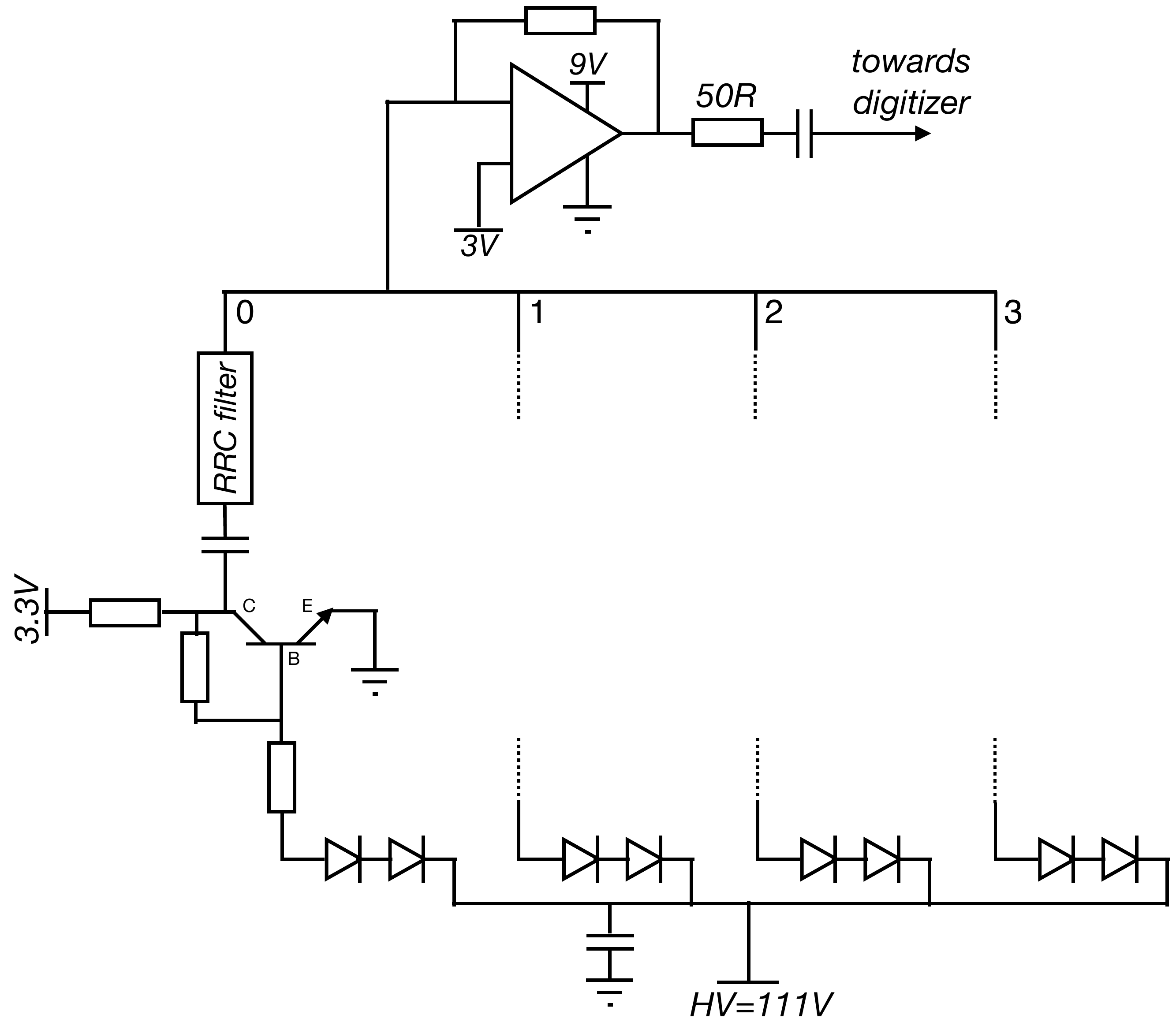}
\caption
{{\it Left:} an amplifier PCB with 8 SiPMs ({\it top}) and its location inside the end-cap when mounted on the bar ({\it bottom}).
{\it Right:} a simplified circuit diagram of the 8\,SiPM amplifier board.
	}
  \label{fig_amplifier2}
\end{figure}

Each scintillator bar is read out on both ends with an array of 8 SiPMs as shown in Fig.\,\ref{fig_amplifier2} (left),
for a total of 236 readout channels.
A plastic end-cap was designed in order to provide stable mounting, easy replacement and 
light-tightness for the SiPM array. 
The inner volume of the end-cap fits the cross-sectional dimension of the SiPM board 
and it bonds the bar as shown in Fig.\,\ref{fig_amplifier2} (left).



Once the assembly of one module was completed,
the module was placed upright, as shown in the photo of Fig.\,\ref{fig_photo} (right), in order to take data with cosmic rays.
The data acquisition module WaveCatcher \cite{Delagnes:WAVECATCHER}, 
two power supplies and the DAQ notebook are also 
shown in the photo.


\section{8 SiPM amplifier board}
\label{sec:amplifier}

The jitter of a timing measurement is inversely proportional to the slope of the pulse leading edge \cite{Delagnes:2016hdo}.
Therefore, to avoid additional dilution of the time resolution, 
it is preferable to keep the rise time of the photosensor response function shorter than
the rise time of the photon emission pulse in the scintillator.
A~large terminal capacitance of SiPM (1.3\,nF for a $6\times6$~mm$^2$ sensor used in this work), however, 
increases the charge depletion time that makes the front edge longer.
It is the main reason why a parallel connection of many sensors
can not be accepted for precision timing measurements.
This issue can be resolved by applying an independent readout for each sensor
in such a way as to isolate the SiPM capacitances from each other. 
The signals are then summed up at the end. 
Another option is to reduce the capacitance  by connecting SiPMs in series 
which decreases the rise time of the leading edge but worsens the signal-to-noise ratio 
\cite{Simon:2018xzl}.
There is an additional time dilution in this case associated with the dependence of transit time on the position of 
SiPM in the series.
Given the above considerations, a combination of parallel and serial connection has been chosen for the amplifier presented here.


Eight surface-mount devices S13360-6050PE\footnote{Specifications: area $6 \times 6$~mm$^2$, pixel pitch 50~$\mu$m, number of pixels 14400, PDE($\lambda=450$\,nm)=40\%, gain $1.7\cdot10^6$, dark counts 2\,Mcpc, crosstalk probability 3\% at $T=25\,^\circ$C and $V_{\rm over}=3V$ \cite{Hamamatsu}.} from Hamamatsu~\cite{Hamamatsu} 
have been soldered in an array to a custom-made PCB.
A schematic of the 8\,SiPM amplifier circuit 
is shown in 
Fig.\,\ref{fig_amplifier2} (right).
The eight SiPMs at each bar end are grouped in four pairs. A series connection is used for SiPMs in a single pair,
while the current of each pair is amplified and shaped independently and 
added to the summing node only at the end.
A positive bias voltage is applied to the SiPM cathode and
the readout line is connected to the anode,
which is close to ground potential.

Taking advantage of the fact that the company provided values of the breakdown voltage
for every individual SiPM, we could sort them in groups by eight having the same properties. 
Thus, each SiPM~array operates using a single bias voltage, which reduces significantly the complexity and 
cost of the board, since fine tuning of individual SiPM is not required.

The input stage of the amplifier is based on the BJT transistor BFP420 operating in the common-emitter configuration.
A photocurrent  generated in the SiPM discharge modifies the base-to-emitter current 
thereby regulating the collector current which, in turn, is converted into voltage before entering the RRC shaper.
The transistor is equipped with negative feedback which makes the circuit more stable, 
reduces distortions of the input waveform and widens the bandwidth 
which justifies a compromise of the gain lowering at this stage.

At the next step, the signal passes the RRC filter.
The filter 
reduces the time constant of the signal trailing edge  
which helps to quickly restore  a stable baseline.
The signal width  is decreased by about 4 times and has a typical value of 10~ns.
The rising edge of the single photon signal was measured to be 2.5~ns.
The power consumption of a single input stage with RRC filter is rather low, 28 mW.
It is important in view of the fact that this part of the circuit is placed right on the back side of the PCB
with respect to SiPMs.
The heat dissipation is carried out by two inner conductive layers of PCB which are grounded and are not used for the circuit.


In order to obtain a combined timing signal, 
four outputs of the shapers are connected together and fed into the output stage 
which is characterised by a  low input impedance.
The trace length from each transistor to the summing node is made equal
to avoid  time smearing due to different transit times.
The output stage of the circuit is represented by a transimpedance amplifier based on a current feedback 
 operational amplifier (OPA) AD8001 which is suitable to this high speed application.
A 50~Ohm resistor at the end makes the output impedance compatible with the input of the digitizer.
The single-ended output signal has a positive polarity.
The RRC shaper and  OPA form a high-pass filter with a 2.9\,MHz low frequency cutoff having a 0.18 gain   and 
a high frequency cutoff of 29\,MHz having a gain of 1.81.
The dynamic range of the amplifier has been adjusted to 1\,V to be  close  to the range of the DAQ digitizer.
The overall power consumption of the circuit is 0.2\,W, which increases the operating board temperature  by almost 4$^\circ$C
over  ambient temperature.

The back side of the board was additionally screened with an aluminium plate which was electrically
connected to the circuit ground. To avoid short-circuits, the plate and board were interleaved with 
a soft thermally conductive interface pad. 


\begin{figure}
\includegraphics[width=0.47\columnwidth]{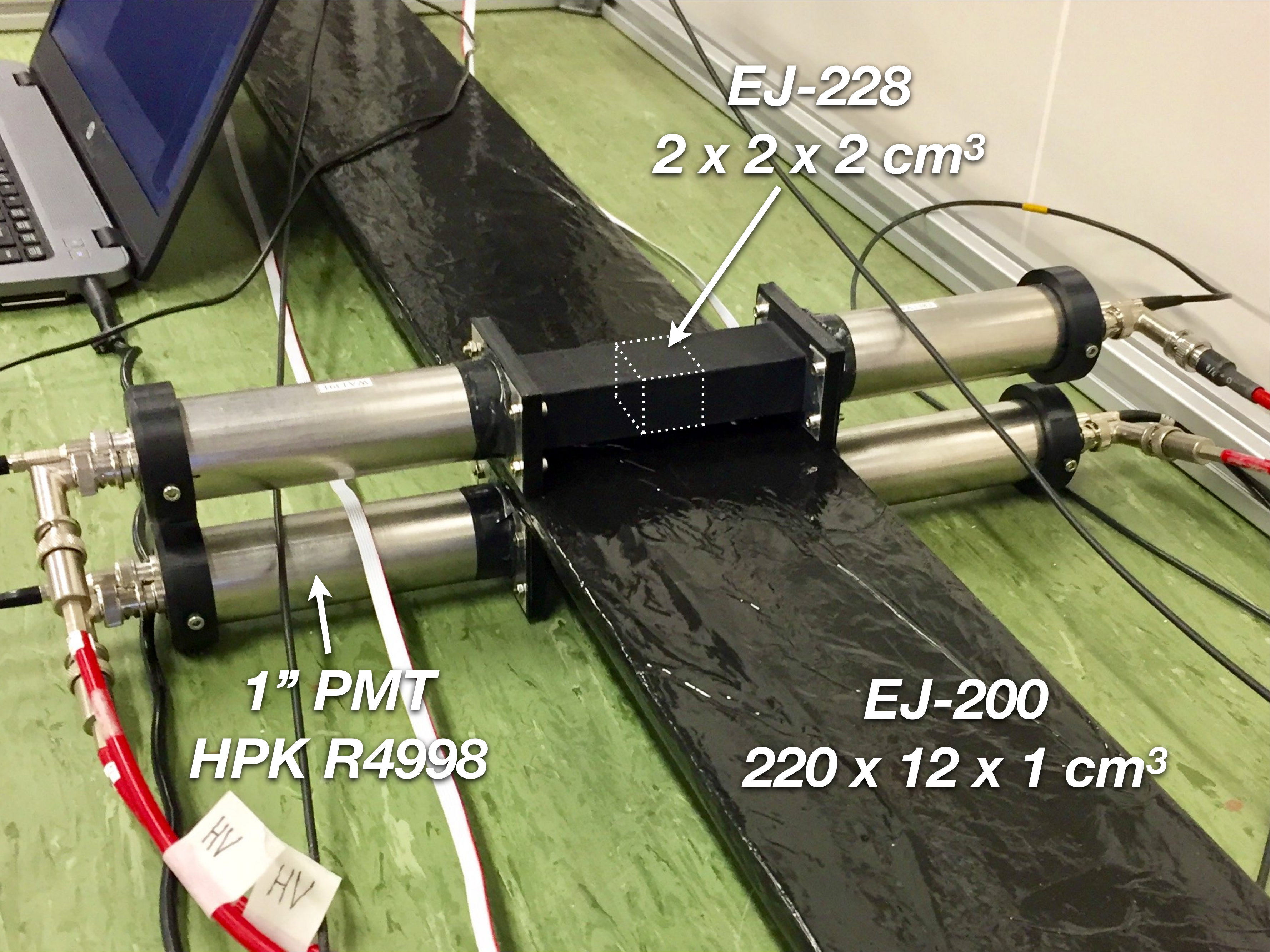}
\hfill
\includegraphics[width=0.49\columnwidth]{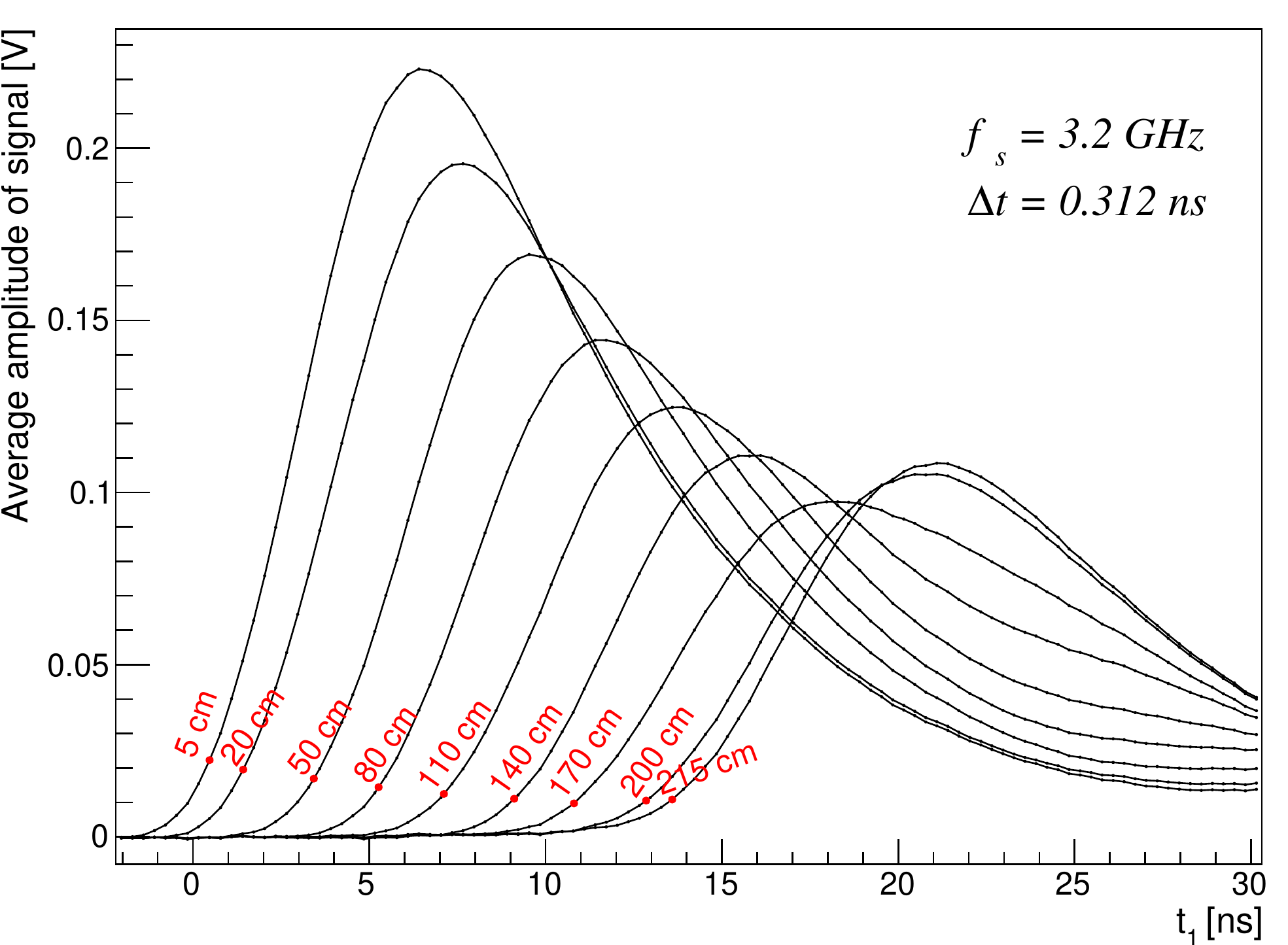}
\caption{{\it Left:} photo of the experimental setup for the cosmic ray measurements showing two reference counters with 
	the bar between them.
	{\it Right:}  averaged waveforms recorded by a single SiPM array.
	The waveforms correspond to 
	different positions of the trigger counters along the bar.
	Markers indicate the time corresponding to 10\% of amplitude as used in the dCFD analysis.
	The corresponding position of trigger counters is also indicated.
	}
  \label{fig_cosmics1}
\end{figure}

\section{Time resolution of a single bar}
\label{sec:aingle_bar}

Characterisation of the time response for a single bar was done using a cosmic test-bench.
The  setup is shown in Fig.\,\ref{fig_cosmics1} (left).
The trigger was formed by the  coincidence of signals of two identical counters located at the top and bottom 
of the bar under test. 
The counters were made of a fast  scintillator, EJ-228, and had a cubic shape with sides of 2~cm. 
They were coupled to 1'' PMTs (Hamamatsu R4998) from two sides via 5~cm long light-guides.
The distance between the centers of the two scintillators was 4~cm.
The trigger time was calculated as the average time registered by the four  PMTs. 


The HV of the trigger PMTs was adjusted in oder to accommodate the signal amplitudes within a 1\,V range.
Only events with amplitudes of all four PMTs larger than 0.5\,V were used in the analysis.
The resolution of the trigger system was obtained from the distribution of 
the time difference measured by top and bottom counters.
The standard deviation $\sigma$ of a Gaussian fit to this distribution was found to be 0.02\,ns.
Since this value is much smaller than the time resolution of the bar,
this uncertainty associated with the trigger counters was neglected in the analysis.

Pulses from four trigger PMTs and two SiPM arrays of the test bar were sent to a WaveCatcher digitizer module \cite{Delagnes:WAVECATCHER} for data acquisition. 
WaveCatcher provides 
 amplitude information from 1024 sampling cells,
thus covering a 320~ns time interval when operating at a  sampling rate of 3.2~GS/s. 
Thus, it gives complete information about the waveform of all pulses.

The measurements were done at 11 positions along the 2.2~m axis of the bar with
an exposure time of  24 hours for each position.
Approximately six hundred events  were taken per exposure. 
For signals from the SiPM arrays, no minimum amplitude cutoff was applied 
so as not to bias the time resolution study.
Not a single event was observed with the trigger fired and not having a signal in the test bar, which indicates the full effectiveness of the counter. 


Pulse waveforms averaged over all events taken for a single exposure and
corresponding to different locations of the trigger counters along the bar are shown in  
Fig.\,\ref{fig_cosmics1} (right).
Due to the photon absorption in plastic, the signal amplitude decreases from 220~mV to 95~mV
for signals corresponding to trigger positions at 5~cm and 170~cm from the photosensors, respectively.
One can also observe a slight increase in the amplitude for trigger positions near the far end of the bar.
It can be explained by backwards reflected light which can provide some improvement in the measurement precision.
A similar effect was observed in Ref.\,\cite{Blondel:2016jju}.
Moreover, the rise time\footnote{Time corresponding to the amplitude interval from 10\% to 90\%.} increases 
from 3.5~ns to 5~ns which is a consequence of the length dispersion of photon paths during their propagation in the bar. 
Both effects, attenuation and  dispersion, lead to a deterioration in the accuracy of measurements.

\begin{figure}[t]
\includegraphics[width=0.5\columnwidth]{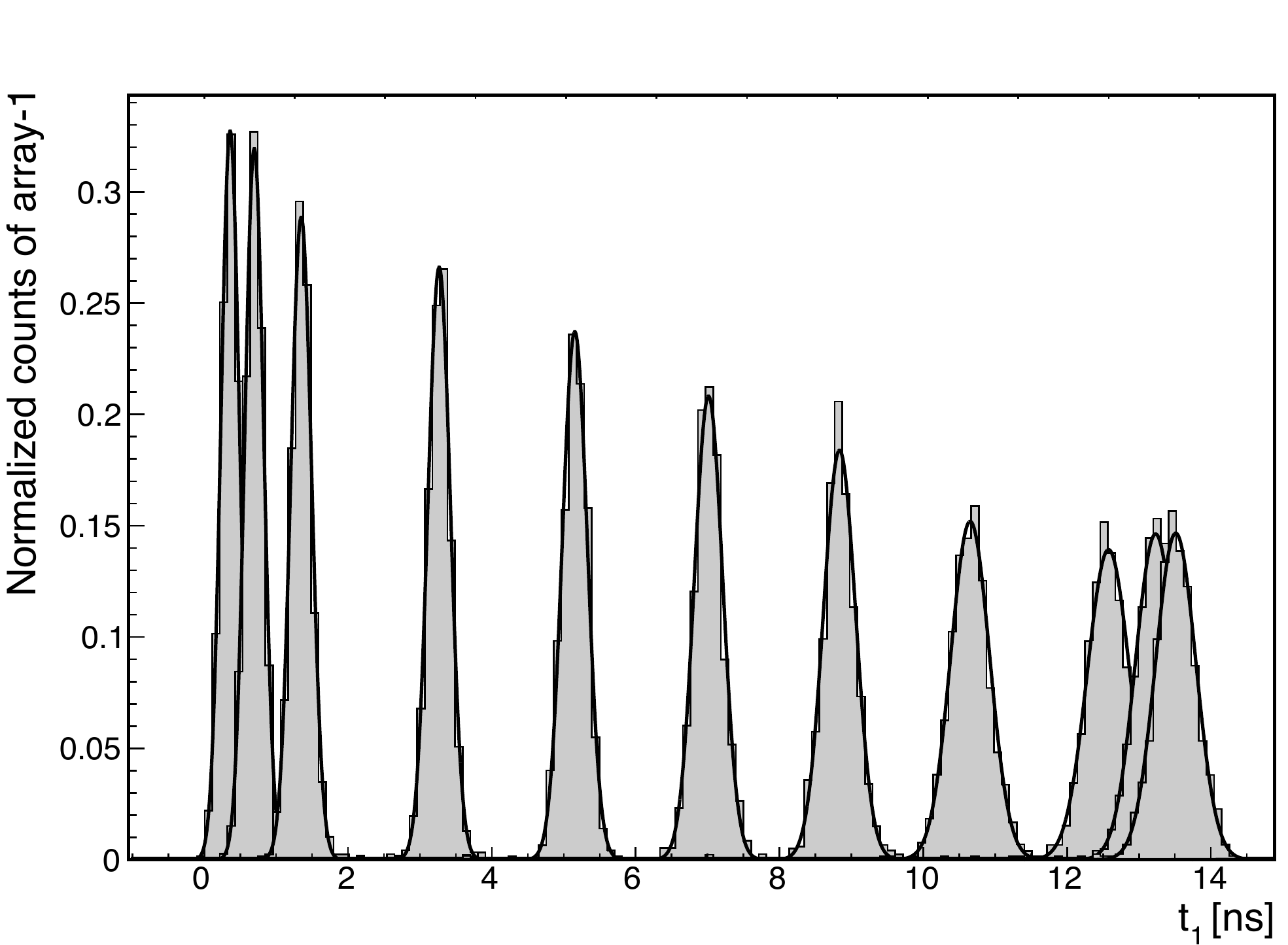}
\hfill
\includegraphics[width=0.47\columnwidth]{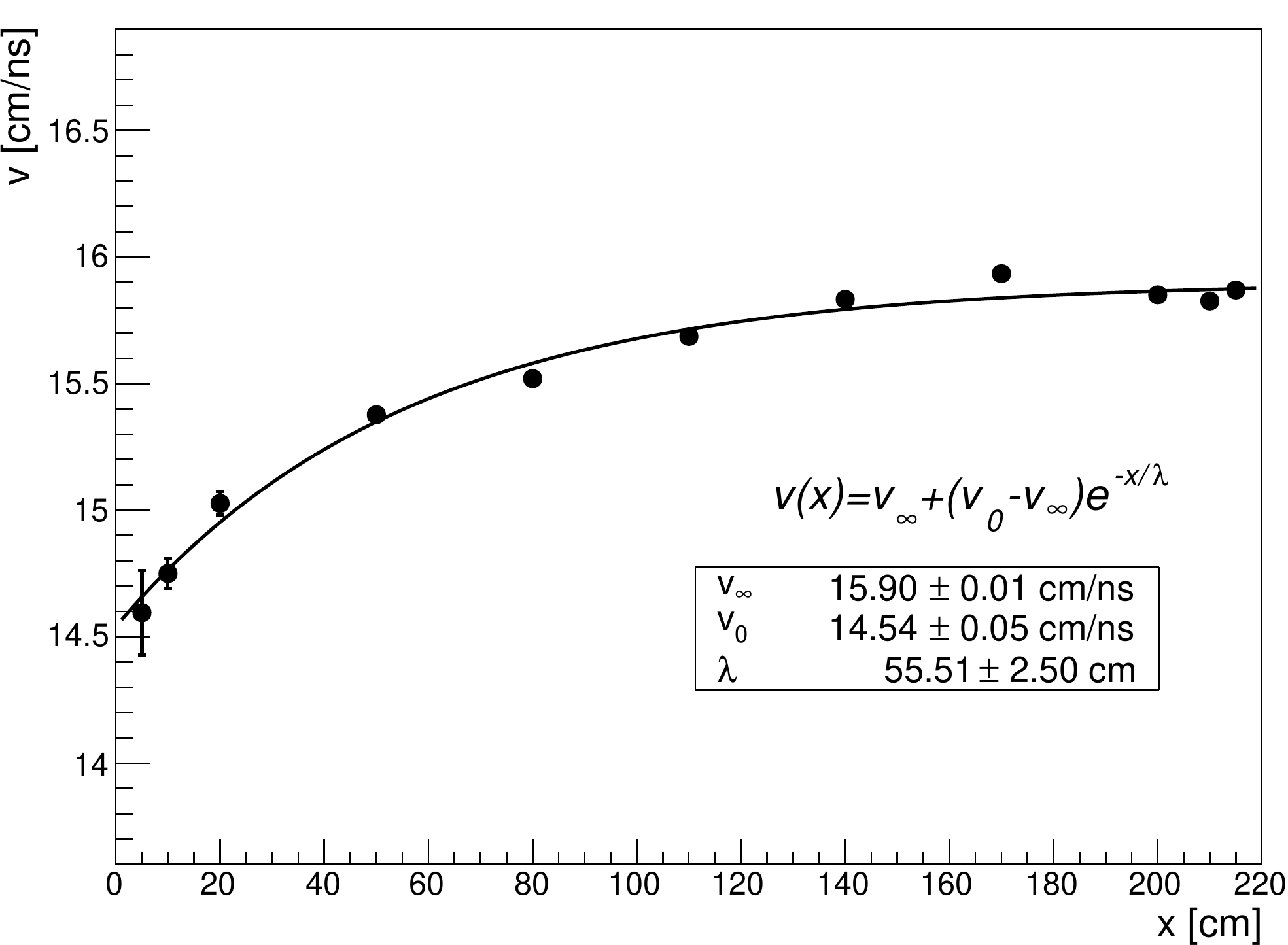}
\caption{{\it Left:}   time detected by the array-1 with respect to the trigger time.
	Different peaks correspond to different positions of the trigger counters along the bar. 
	{\it Right:} an effective light velocity versus position of the trigger counters along the bar.
	}
  \label{fig_peaks1}
\end{figure}

A digital constant fraction discrimination (dCFD) technique was used to define the arrival time of signals~\cite{Delagnes:2016hdo}.
The  time corresponding to 10\% of the signal amplitude was taken in the analysis. 
 A linear interpolation between amplitudes of adjacent time cells was applied.
In the following discussion, the two SiPM arrays at the two ends of the bar are referred to as array-1 and array-2.
The difference between the time detected by array-1
and the trigger time is shown in Fig.\,\ref{fig_peaks1} (left) 
for different positions of the trigger counters along the bar.
The standard deviation of a Gaussian approximation to this distribution defines the time resolution of the measurement,
while  the mean value  defines the arrival time.
In general, the dependence of the propagation distance on the arrival time  is close to linear.
The ratio of  these values represents the effective light propagation velocity 
which is shown in 
Fig.\,\ref{fig_peaks1} (right).
A fit to the distribution by the sum of a constant and exponent shows that
the velocity increases from 14.5~cm/ns to 15.9~cm/ns with distance. This is consistent with the fact 
that photons at larger angles with respect to the surface of the bar (slower velocity) have to travel longer distances 
and are more prone to attenuation.
The interaction time $t_{int}$ is calculated using the time registered by the arrays at both ends of the bar $t_1$ and $t_2$
and the  corresponding velocities  $v_1$ and $v_2$
\begin{equation}
t_{int} = \frac{t_1 v_1 + t_2 v_2}{v_1 + v_2} - \frac{L}{v_1 + v_2}  
\stackrel{\,v_1=v_2\,}{=}
\frac{t_1 + t_2}{2} - \frac{L}{2v}   ,
\label{Eq1}
\end{equation}
where $L$ is the bar length.
An assumption of a constant propagation velocity, which is commonly used in analyses, significantly simplifies the formula.
Using the velocity parametrization shown in Fig.\,\ref{fig_peaks1} (right) one can determine  the bias introduced by
this assumption. It does not exceed 0.05~ns which is smaller by a factor of about four than the required resolution.

\begin{figure}[t]
\includegraphics[width=0.49\columnwidth]{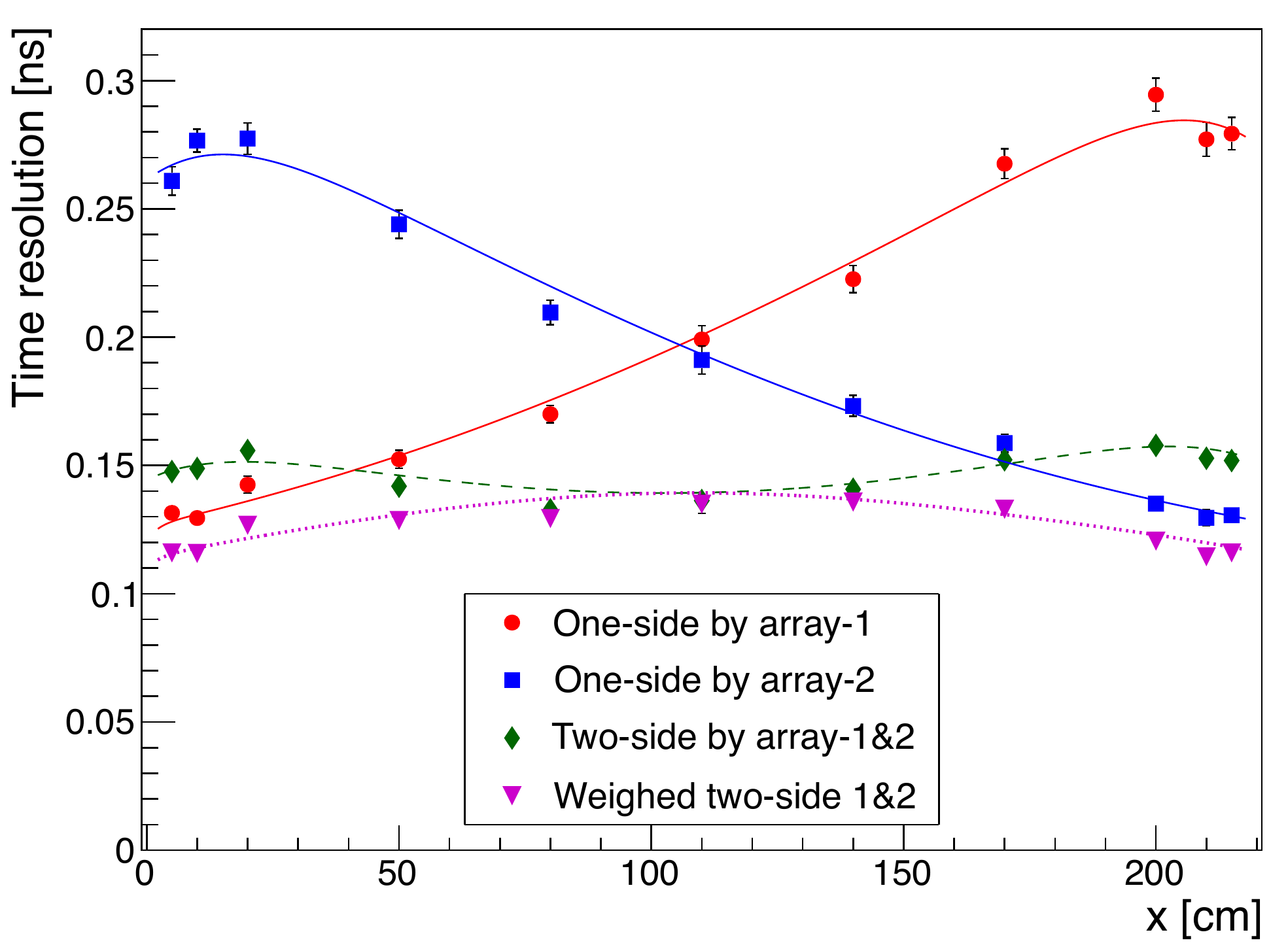}
\hfill
\includegraphics[width=0.49\columnwidth]{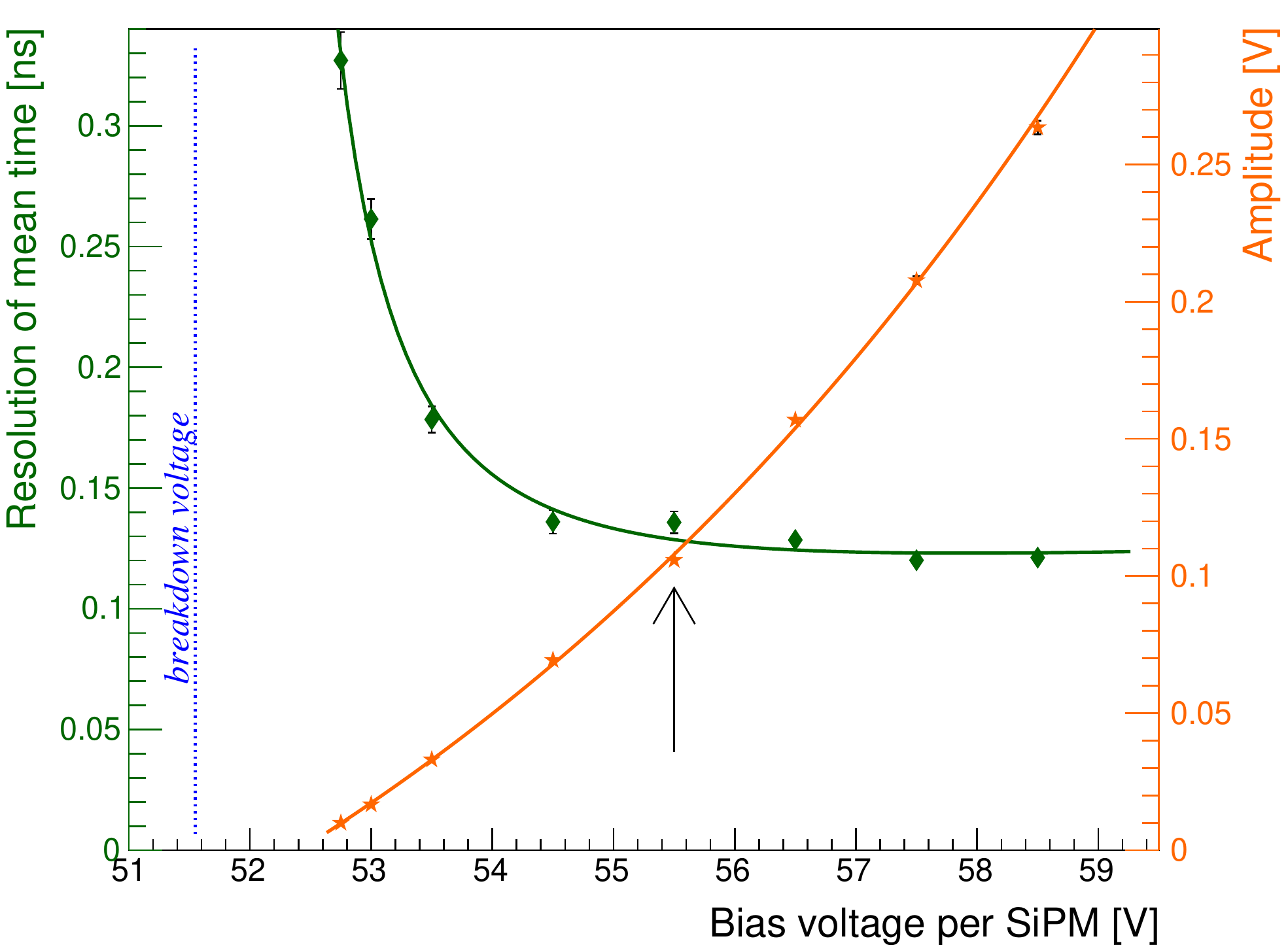}
\caption{{\it Left:}  time resolution of one-side readout (approximated by solid curves), the resolution of two-side readout
(approximated by broken curves) as a function of distance along the bar. 
{\it Right:} two-side resolution (left axis of histogram) and the signal amplitude (right axis of histogram) at the center of the bar, 
	$x = 110$~cm, versus the bias voltage.  
	The voltage used during datataking is indicated by the arrow.
}
  \label{fig_sigma1}
\end{figure}

The time resolution as a function of position of the interaction point along the bar 
obtained with one-side measurements using array-1 or array-2  is shown in Fig.\,\ref{fig_sigma1} (left). 
The graphs  were approximated by 
analytic functions consisting of a sum of two exponential functions and a constant.
The resolution for each array on either end  degrades from 0.13~ns to 0.27~ns with distance.
The resolution of two-side readout was obtained from the gaussian approximation of the $t_{int}$ distribution
calculated with Eq.\,\ref{Eq1} and it is also shown in Fig.\,\ref{fig_sigma1} (left).
Even better precision for  $t_{int}$  can be achieved by weighting $t_1$ and $t_2$ 
for each event with their respective resolutions. 
Unlike the case of one-side readout, 
the two-side graphs are almost independent of the interaction position and show a time resolution of 0.14~ns at the center of the bar.

The dependence of the time resolution and the most probable value of the amplitude 
versus the SiPM bias voltage for interactions at the bar center is shown in 
Fig.\,\ref{fig_sigma1} (right). 
As the photon detection efficiency (PDE) improves with increasing voltage,
the values of amplitude increase and the resolution improves.
The latter however is practically saturated at  overvoltage\footnote{The difference between the bias and breakdown voltages.} 
values above 4\,V. 
This value corresponds to a  bias voltage of 111\,V applied to a SiPM pair in series.
This voltage was used during the datataking and it is indicated in the figure by an arrow.


\section{Conclusions}
\label{sec:conclusions}

A time-of-flight system designed for the upgrade of the ND280 detector has been presented.
An array of  plastic scintillator bars is used as an active element of the detector.
A new approach for the light readout, which employs arrays of large-area SiPMs at two ends of each bar,
has been dictated by the limited space inside the basket of ND280 and the requirement of high time resolution.
Six modules of ToF have been successfully assembled, placed upright in the temporary support frame and tested with cosmics.
The time response of one bar was characterised 
as a function of position along the bar
and the time resolution was found to be 0.14~ns on average,
which meets the requirements of the ND280 upgrade.
The ToF modules will be packed without disassembly and shipped from CERN to J-PARC. 
They will be installed in the basket of ND280 and will provide almost $4\pi$ coverage for the active neutrino target and two TPCs.
The modules will be used to veto charged particles originated outside the neutrino target
and, potentially, for  particle identification starting from the year 2023.


\section*{Acknowledgements}

The construction of the ToF detector was made possible thanks to funding from 
the Swiss National Science Foundation under Grant FLARE S19516.
The R\&D part of the project was covered by the  grant PP00P2\_150583.
The Russian group acknowledges the support from
JSPS-RFBR grant \#20-52-50010, Russia, and  the support in
the framework of the State project "Science" by the Ministry of Science
and Higher Education of the Russian Federation under the contract
075-15-2020-778.
T.\,Lux acknowledges funding from the Spanish Ministerio de Econom\'{i}a y Competitividad (SEIDI - MINECO) under Grants No.~PID2019-107564GB-I00 and SEV-2016-0588. IFAE is partially funded by the CERCA program of the Generalitat de Catalunya.
We also would like to acknowledge the contribution of FAST (COST action TD1401) which was at the origin of this project.


\bibliographystyle{JHEP}
\bibliography{paper}

\end{document}